\documentclass[reprint,twocolumn,amsmath,amssymb,aps,prl,superscriptaddress]{revtex4-1}
\usepackage{graphicx}
\usepackage{dcolumn}
\usepackage{bm}
\usepackage{float}
\usepackage{amssymb}
\usepackage{color}
\usepackage{multirow}
\usepackage{stmaryrd}
\begin{document}

\title{Crystal growth and quantum oscillations in the topological chiral semimetal CoSi}

\author{Xitong Xu}
\affiliation{International Center for Quantum Materials, School of Physics, Peking University, China}

\author{Xirui Wang}
\affiliation{International Center for Quantum Materials, School of Physics, Peking University, China}

\author{Tyler A. Cochran}
\affiliation{Laboratory for Topological Quantum Matter and Spectroscopy (B7), Department of Physics, Princeton University, Princeton, New Jersey 08544, USA}

\author{Daniel S. Sanchez}
\affiliation{Laboratory for Topological Quantum Matter and Spectroscopy (B7), Department of Physics, Princeton University, Princeton, New Jersey 08544, USA}

\author{Ilya Belopolski}
\affiliation{Laboratory for Topological Quantum Matter and Spectroscopy (B7), Department of Physics, Princeton University, Princeton, New Jersey 08544, USA}

\author{Guangqiang Wang}
\affiliation{International Center for Quantum Materials, School of Physics, Peking University, China}

\author{Yiyuan Liu}
\affiliation{International Center for Quantum Materials, School of Physics, Peking University, China}

\author{Hung-Ju Tien}
\affiliation{Department of Physics, National Cheng Kung University, Tainan 701, Taiwan}

\author{Xin Gui}
\affiliation{Department of Chemistry, Louisiana State University, Baton Rouge, LA 70803, USA}

\author{Weiwei Xie}
\affiliation{Department of Chemistry, Louisiana State University, Baton Rouge, LA 70803, USA}

\author{M. Zahid Hasan}
\affiliation{Laboratory for Topological Quantum Matter and Spectroscopy (B7), Department of Physics, Princeton University, Princeton, New Jersey 08544, USA}
\affiliation{Lawrence Berkeley National Laboratory, Berkeley, California 94720, USA}

\author{Tay-Rong Chang}
\affiliation{Department of Physics, National Cheng Kung University, Tainan 701, Taiwan}
\affiliation{Center for Quantum Frontiers of Research and Technology (QFort), Tainan 701, Taiwan}

\author{Shuang Jia}
\email{gwljiashuang@pku.edu.cn}
\affiliation{International Center for Quantum Materials, School of Physics, Peking University, China}
\affiliation{Collaborative Innovation Center of Quantum Matter, Beijing 100871, China}
\affiliation{CAS Center for Excellence in Topological Quantum Computation, University of Chinese Academy of Sciences, Beijing 100190, China}
\affiliation{Beijing Academy of Quantum Information Sciences, West Building 3, No. 10 Xibeiwang East Road, Haidian District, Beijing 100193,China}

\begin{abstract}
We survey the electrical transport properties of the single-crystalline, topological chiral semimetal CoSi which was grown via different methods.
High-quality CoSi single crystals were found in the growth from tellurium solution.
The sample's high carrier mobility enables us to observe, for the first time, quantum oscillations (QOs) in its thermoelectrical signals.
Our analysis of QOs reveals two spherical Fermi surfaces around the $R$ point in the Brillouin zone corner. The extracted Berry phases of these electron orbits are consistent with the $-2$ chiral charge as reported in DFT calculations.
Detailed analysis on the QOs reveals that the spin-orbit coupling induced band-splitting is less than 2~meV near the Fermi level, one order of magnitude smaller than our DFT calculation result.
We also report the phonon-drag induced large Nernst effect in CoSi at intermediate temperatures.

\end{abstract}

\pacs{}
\date{\today}
\maketitle

\section{Introduction}

Topological materials have been a hot research frontier since the discovery of topological insulators and semimetals~\cite{RevModPhys.82.3045,RevModPhys.83.1057,ando2013topological,burkov2016topological}. Fundamental fermionic particles in high-energy physics, including the Majorana, Dirac and Weyl fermions~\cite{1.3549729}, exist in momentum space as low-energy quasi-particle excitations in these materials' electronic bands. One example is the Weyl semimetal which features linear crossings of two non-degenerate bands carrying Chern number  $\left|C\right|=1$~\cite{burkov2016topological,jia2016weyl,YanTopological,BurkovWeyl}. These band crossings, dubbed as spin-1/2 Weyl nodes, necessitate the presence of exotic helicoid surface states and Fermi arcs. Recent theories in condensed matter systems have expanded massless fermions with higher quantized Chern numbers which have no counterparts in elementary particles. These ``new fermions''~\cite{Bradlynaaf5037} include the spin-1 chiral fermion~\cite{PhysRevB.85.155118,PhysRevLett.119.206402}, double Weyl fermion~\cite{PhysRevLett.119.206402,PhysRevA.94.053619}, spin-3/2 Rarita-Schwinger-Weyl fermion~\cite{PhysRev.60.61,PhysRevB.93.045113,PhysRevB.94.195205}, etc.

Cobalt monosilicide crystalizes in a chiral structure in the P2$_1$3 space group (Fig.~\ref{f1}(a) inset).
It has been intensely studied as a potential thermoelectric material~\cite{PhysRev.134.A774,fedorov1995semimetals,kim2002effect,lue2004hole,ren2005effects,skoug2009high} due to its large power factor at room temperature. Recent theoretical works have found that CoSi and its isostructural siblings possess a chiral double sixfold-degenerate spin-1 Weyl fermion at $R$ point and a fourfold-degenerate chiral fermion at $\Gamma$ point in their Brillouin zone (BZ)~\cite{PhysRevLett206401,PhysRevLett206402,pshenay2018band}. These ``new fermions'' with large topological charges are connected by long, robust Fermi arcs on the surface which have been later confirmed by angle-resolved photoemission spectroscopy (ARPES) experiments~\cite{sanchez2018discovery,PhysRevLett.122.076402,rao2019new}.
However, no transport properties of CoSi directly related to its topological nature have been reported until now.
As far as we are aware, previous reported CoSi single crystals manifest relatively low carrier mobilities and no quantum oscillations (QOs) have been observed in their electrical properties under magnetic field. It is difficult to bridge the transport properties with its topological band structure due to low sample quality.

Here we report our crystal growth and a survey of the electrical transport properties of single-crystalline CoSi. Tellurium was found to be an appropriate metal flux for the growth of CoSi which yields high-quality single crystals with large magneto-resistance (MR) and carrier mobilities.
Although there is plenty of research work on the thermopower of CoSi~\cite{PhysRev.134.A774,fedorov1995semimetals,kim2002effect,lue2004hole,ren2005effects,skoug2009high,sun2013effects,sun2016isovalent,burkov2017transport,pshenay2018electronic,pshenay2018effect},
few of them have paid attention to its magneto-Seebeck and Nernst effect.
Combining the high quality of our Te-flux grown samples and the sensitivity of thermoelectricity measurement~\cite{behnia2015fundamentals}, we are able to observe, for the first time, QOs in the thermoelectric signals of CoSi.
By analyzing the QOs in magneto-Seebeck and Nernst signals at different temperatures and magnetic field directions, we reveal two spherical Fermi surfaces around the BZ corner $R$ point, which is consistent with our density functional theory (DFT) calculations.
The extracted Berry phases of electron orbit equal zero, agreeing well with the scenario of $-2$ chiral charge at $R$ point.
We also found that the spin-orbit coupling (SOC) induced band-splitting is less than 2~meV near the Fermi level and this result is one order smaller than our DFT calculations.
We also report a large Nernst effect in CoSi due to the combination of high mobility and phonon-drag contribution at intermediate temperatures. As a consequence, a relatively large Nernst-Ettinshausen figure of merit (ZT$_\epsilon$)~\cite{nolas2013thermoelectrics} of around 0.03 is achieved at 42~K in 14~T.

\section{Method}

\begin{table*}[htbp]
\footnotesize
\caption{\label{t1}
Summary of the electrical transport properties of the CoSi single crystals from different growth conditions. Electric current is applied in the basal plane. $\mu_e$ and n$_e$ are the estimated mobility and concentration of electrons at 2~K.
}
\begin{tabular}{p{1.5cm}<{\centering}p{1.8cm}<{\centering}p{1.5cm}<{\centering}p{2cm}<{\centering}p{2cm}<{\centering}p{2cm}<{\centering}p{6cm}<{\centering}}
\hline\hline
Sample      &   Basal Plane &   RRR     &   MR$_{9T}$ (\%)  &   $\mu_e$($cm^2/Vs$)  &   n$_e$($10^{20}cm^{-3}$) &   Growth Condition    \\
\hline
F01         &   --          &   3.33    &   5               &   640                 &   1.92                    &   Floating zone       \\
\hline
I01         &   (111)       &   1.63    &   -1.4            &   280                 &   1.05                    &   \multirow{3}*{CVT, Co+Si+0.1g I$_2$, 900 to 1100$^{\circ}$C}\\
\cline{1-6}
I02         &   (111)       &   1.67    &   -1.2            &   350                 &   1.29                    &                       \\
\cline{1-6}
I04         &   (110)       &   7.33    &   29              &   1300                &   1.49                    &                       \\
\hline
I03         &   --          &   2.56    &   1.7             &   410                 &   2.42                    &   CVT, Co+Si+5mg I$_2$, 900 to 1000$^{\circ}$C\\
\hline
I05         &   (100)       &   8.32    &   34.9            &   2050                &   2.87                    &   CVT, CoSi+5mg I$_2$, 1000 to 1100$^{\circ}$C\\
\hline
B01         &   (111)       &   2.87    &   -0.4            &   --                  &   --                      &   \multirow{2}*{CVT, Co+Si+1.5mL Br$_2$, 1100 to 900$^{\circ}$C}\\
\cline{1-6}
B02         &   (110)       &   1.66    &   -4              &   --                  &   --                      &                       \\
\hline
S01         &   --          &   1.61    &   4.8             &   --                  &   --                      &   Flux, CoSi+19Sb, 1300 to 750$^{\circ}$C\\
\hline
S02         &   --          &   1.97    &   9.8             &   230                 &   3.02                    &   Flux, CoSi+19Sb, 1050 to 700$^{\circ}$C\\
\hline
T01         &   (111)       &   19.9    &   314             &   7300                &   2.55                    &   \multirow{2}*{Flux, Co+Si+20Te, 1050 to 700$^{\circ}$C}\\
\cline{1-6}
T02         &   (111)       &   29.5    &   398             &   6300                &   1.47                    &                       \\
\hline\hline

\end{tabular}
\end{table*}

We used floating zone, chemical vapor transport (CVT) and flux method to grow CoSi single crystals successfully. In a floating zone growth, polycrystalline CoSi was firstly prepared by arc-melting stoichiometric cobalt and silicon chunks and then the crystal was grown in a Quantum Design IR image furnace. In a standard CVT method, either CoSi powder or fresh cobalt and silicon powder was used as starting materials. About 1 gram raw material was loaded into a $\Phi 10 \times 22\ \mathrm{mm}$ fused silica tube together with transport agent I$_2$ or Br$_2$. The tube was then sealed in vacuum and the transport reaction was maintained at high temperatures for around 10 days. Shiny, millimeter-sized polyhedron-like crystals were yielded. In the growth from liquid solution, antimony, tin and tellurium were chosen as the flux. In a growth from tellurium flux, cobalt, silicon and tellurium powder of the molar ratio of 1:1:20 were set in an alumina crucible and then sealed in a fused silica ampoule in vacuum. After soaking at 1050$^{\circ}$C for a few hours, the ampule was slowly cooled down to the centrifuging temperature 700$^{\circ}$C at the rate of 3$^{\circ}$C/h. Several single crystalline chunks developing large (111) facets were obtained. Similar procedure was carried out in antimony and tin flux but the resulting crystals were smaller and rod-like. Detailed information is summarized in Table~\ref{t1}.

Electrical properties of as-obtained CoSi crystals were mainly characterized in a Quantum Design physical property measurement system (PPMS-9).
Thermoelectric measurement was carried out in a 14 T Oxford Teslatron PT system, using a one-heater-two-thermometer setup in which the temperature gradient is applied within the crystallographic (111) plane and magnetic field $\bm{H}$ is applied along the $\left[ 111 \right]$ direction. Vacuum better than $3\times 10^{-5}\, \mathrm{Pa}$ was maintained during the measurement.
The thermoelectric voltage signals were amplified by using EM DC Amplifier A10 and subsequently collected in a Keithley 2182A nano-voltmeter.
Angle-dependent Nernst signals were measured in a rotating probe in the same magnet in which the vacuum was kept at around $10^{-3}\, \mathrm{Pa}$  for effective heat-sink of the probe.
ARPES measurements were carried out at the ADRESS beamline of the Swiss Light Source using a SPECS PHOIBIOS 150 analyzer. Natural facets of CoSi single crystals were Ar sputtered and annealed \it in situ \rm in ultra-high vacuum before being measured at 13 K. Spectra were acquired using 550~eV circular right and circular left incident photons with an energy resolution of 85~meV and angular resolution $<0.07$ degrees. Right and left circular spectra were summed to eliminate the significant effect of circular dichroism.
Band structure calculations were performed under the framework of the generalized gradient approximation of DFT~\cite{PhysRevLett.77.3865} as implemented in the VASP package~\cite{kresse1996efficiency}.

\section{Experimental Results and Discussions}

\begin{figure*}[htbp]
\begin{center}
\includegraphics[clip, width=0.9\textwidth]{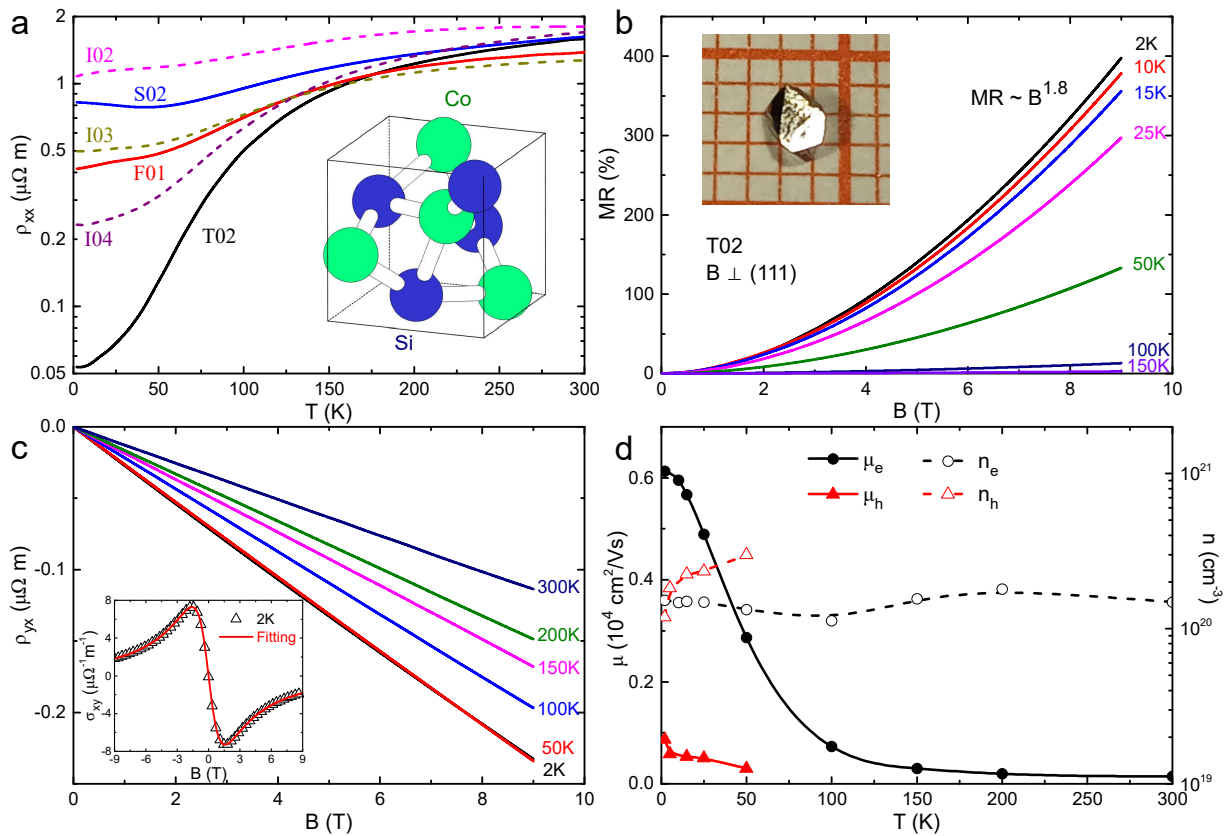}\\[1pt]  
\caption{Color online. (a) $\rho (T)$ curves of selected CoSi crystals with RRR ranging from $1.7$ to around $30$ (see Table~\ref{t1}). Note the log scale of $\rho _{xx}$. Inset: Unit cell of the chiral crystal.   (b) MR of sample T02. Inset: A photo of the Te-flux-grown CoSi revealing shiny (111) facets.   (c) Hall resistivity of T02. Inset: Two-band fitting of the Hall conductivity.   (d) Carrier mobilities and concentrations in T02 versus temperature.}
\label{f1}
\end{center}
\end{figure*}

Room temperature resistivity of our CoSi samples generally ranges from $1\sim2\ \mu\Omega\,m$, close to previously reported values for both poly- and single-crystalline CoSi~\cite{skoug2009high,petrova2010elastic,stishov2012self,samatham2014investigations,burkov2017transport}.
All the samples show metallic temperature-dependent resistivity ($\rho (T)$) in Fig.~\ref{f1}(a), but their residual resistance ratios (RRR, defined as the ratio of R(300K)/R(2K)) differ from each other. We notice that the $\rho (T)$ curves of the crystals grown via CVT method have apparent sample dependence, e.g., the RRR of sample I02 and I04 which were grown in the same condition differed by four times (See Table~\ref{t1}).
Moreover, the  $\rho (T)$ curves of the samples with small RRR tends to have an upturn feature below 50~K while their MR at 9~T is negative and small (Table~\ref{t1}).
These features are similar to what is reported in Ref.~\cite{stishov2012self}, suggesting large magnetic defect concentrations in the samples with small RRR.

In this study we focus on the single crystals grown from tellurium flux.
These crystals are about 2~mm in size and prone to develop large (111) planes.
No residual Te was found within the accuracy of energy dispersive spectroscopy.
These crystals have the largest RRR ($\sim 30$) ever reported to our knowledge, and hereafter we focus on sample T02.

The quasi-quadratic profile of MR ($\propto B^{1.8}$) for sample T02 (Fig.~\ref{f1}(b)) is similar to what was observed in arc-melted samples~\cite{samatham2014investigations} but the value is larger ($\mathrm{MR}\!=\!400\%$ at 2~K in 9~T).
Yet the MR is still smaller than those of typical Dirac and Weyl semimetals~\cite{he2014quantum,PhysRevB.95.085202,zhang2017magnetic,ali2014large}, indicating relatively low carrier mobilities.
The Hall resistivity measurement shows a negative, nearly linear field dependence (Fig.~\ref{f1}(c)) and the slope remains unchanged below 50~K. We fit the off-diagonal conductivity tensor by using a two-band model~\cite{hurd2012hall}, $$\sigma_{xy}=\left\lbrack n_e\mu_e^2 \frac{1}{1+(\mu_eB)^2}-n_h\mu_h^2 \frac{1}{1+(\mu_hB)^2}\right\rbrack eB$$
where $n_e$($n_h$) and $\mu_e$($\mu_h$) denote the carrier concentration and mobility of electrons (holes), and $e=-|e|$ is the electron charge hereafter for clarity.
Above 50 K, a simple one-carrier model is used to estimate the concentration and mobility of electrons because of the small MR. As shown in Fig.~\ref{f1}(d), the carrier concentration of both electrons and holes is around $1.5\times 10^{20}\,\mathrm{cm^{-3}}$ at 2~K, while the mobility of electrons ($\sim 6000\ \mathrm{cm^2/Vs}$) is much larger than that of holes. This is reasonable due to the fact that hole carriers in CoSi are much heavier than electrons~\cite{PhysRev.134.A774,pshenay2018electronic}.
The electron concentration remains intact in the whole temperature range while the mobility drops abruptly above 50~K, reflecting significant reduction of mean free path of the carriers at higher temperature.

\begin{figure*}[htbp]
\begin{center}
\includegraphics[clip, width=0.9\textwidth]{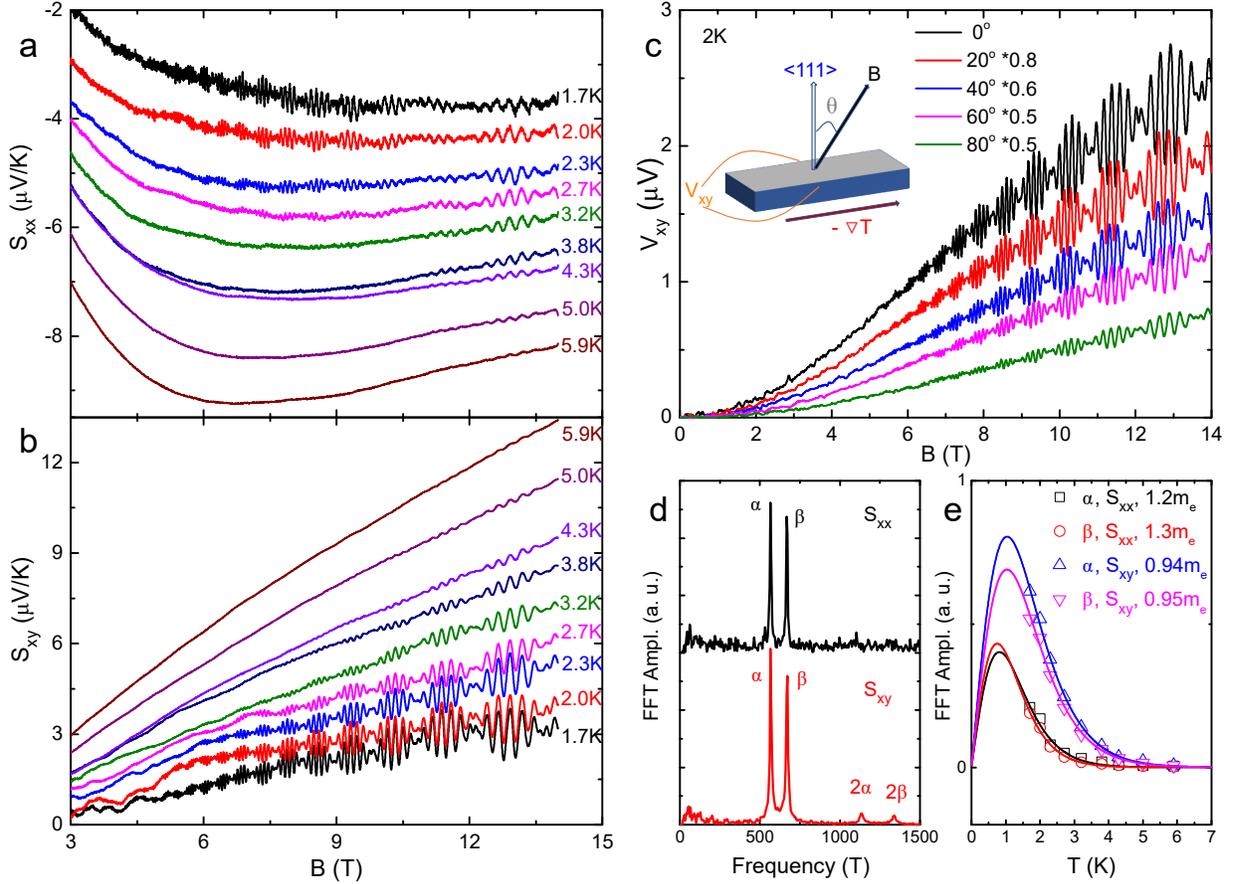}\\[1pt]  
\caption{Color online. (a) and (b) Magneto-thermopower S$_{xx}$ and S$_{xy}$ at selected temperatures, showing strong QOs. Magnetic field is along the crystallographic $\left[ 111 \right]$ orientation. (c) Nernst voltages at selected angles. Magnetic field is tilted within the plane expanded by the $\left[ 111 \right]$ direction and the direction of the temperature gradient. Note that the voltages have been multiplied by some coefficients for clarity.    (d) FFT spectra of S$_{xx}$ and S$_{xy}$. (e) Fitting of cyclotron mass $m^\ast$ using the derivative of the LK formula.}
\label{f2}
\end{center}
\end{figure*}

Figure~\ref{f2}(a) and (b) show $S_{xx}$ and $S_{xy}$ of sample T02 at low temperatures. Strong QOs with apparent beating pattern are observed in both $S_{xx}$ and $S_{xy}$ at base temperature, which indicates that there exist two oscillatory parts close in magnitude and frequency. Though the QOs quickly damp with increasing temperature, this beating feature and relative phase remain unchanged.
Our fast Fourier transformation (FFT) reveals two fundamental frequencies of 568~T and 671~T in Fig.~\ref{f2}(d), labelled as $\alpha$ and $\beta$ orbits, respectively.

To extract the cyclotron masses of the orbits, we analyze the QOs at different temperatures.
The temperature dependence of the QOs in resistivity and magnetization is well described by the Lifshitz-Kosevich (LK) formula~\cite{shoenberg2009magnetic} as following:
$$R_T=\frac{\alpha p X}{\sinh \alpha p X}=D(\mathbb{X})$$
where $\alpha=2\pi^2 k_B/e\hbar$, $X=m^\ast T/B$, $m^\ast$ being the cyclotron mass, and $\mathbb{X}=\alpha p X$.
Previous works pioneered by Fletcher \it et al.\rm  ~\cite{fletcher1981amplitude,fletcher1983experimental,coleridge1989low,fletcher1995oscillations,tieke1996magnetothermoelectric, morales2016thermoelectric} suggest that the thermal damping factor for diffusive part of magneto-thermopower should be
$$R_T=|D'(\mathbb{X})|=\frac{(\alpha pX)\coth(\alpha pX)-1}{\sinh (\alpha pX)}$$
as the QOs in $S_{ij}$ depend on the derivative of density of states. N. B. there is no apparent contribution to QOs from phonon-drag below 10~K because the carrier density of CoSi is high and the Fermi surface is large. Moreover there is no drift of the oscillatory phases at different temperatures.
The cyclotron masses of $\alpha$ and $\beta$ are fitted to be 1.2 $m_e$ and 1.3 $m_e$ in $S_{xx}$  and 0.94 and 0.95 in $S_{xy}$, respectively. These cyclotron masses are much heavier than those of electrons observed in Weyl semimetal TaAs and TaP family~\cite{PhysRevB.95.085202,zhang2016signature,zhang2017magnetic}, but close to previously reported effective masses of electrons ($\sim2\, m_e$) in CoSi~\cite{PhysRev.134.A774,pshenay2018electronic}.
The difference between cyclotron masses obtained from $S_{xx}$ and $S_{xy}$ deserves better understanding in the future.

In order to map the Fermi surface in the momentum space, we also performed Nernst voltage measurement when the magnetic field is along different orientations. The field is laid in the plane expanded by crystallographic $\left[ 111 \right]$ direction and $-\nabla T$ (Fig.~\ref{f2}(c)). Aside from the relative magnitude, the oscillatory pattern of the Nernst voltages remains almost unchanged over the whole $2\pi$ angle. This result suggests that the two Fermi surfaces involved are spherical.

\begin{figure*}[htbp]
\begin{center}
\includegraphics[clip, width=0.9\textwidth]{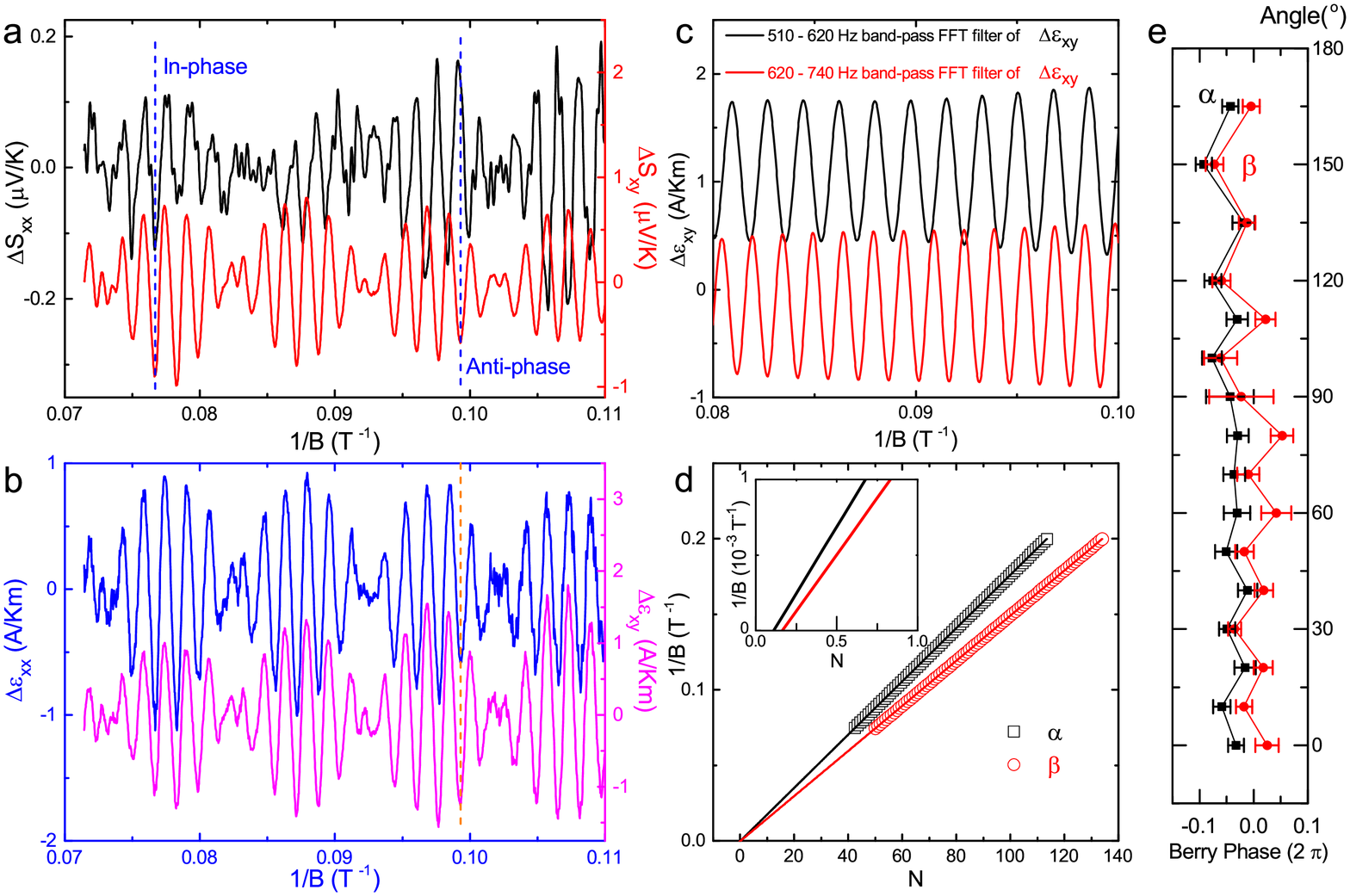}\\[1pt]  
\caption{Color online. (a) Oscillatory part of S$_{xx}$ and S$_{xy}$ at 1.7~K. Note different scales.     (b) Oscillatory part of the thermoelectric tensor $\epsilon_{xx}$ and $\epsilon_{xy}$.    (c) $\Delta\epsilon_{xy}$ after passing band-pass FFT filter.    (d) Landau fan diagram. Peak positions in (c) are assigned with integral values while valley positions with half-integrals. Inset: Zoom-in near N = 0.   (e) Angle-dependence of extracted Berry phases from V$_{xy}$. The error-bar is set to be 3 times the value of standard error.}
\label{f3}
\end{center}
\end{figure*}

Before we proceed the analysis on the Berry phases of these two orbits, the criterion for numbering the Landau index has to be clarified in thermoelectric measurements.
Previous study selected the peak positions of the oscillatory part of $S_{xx}$ as integral Landau indices (N. B. here the integral is corresponding to the Fermi level leaving the N-th Landau level) and further included the additional phase from thermoelectrical measurement~\cite{matusiak2017thermoelectric}. The criterion for detecting the Landau indices in $S_{xy}$ is more complicated~\cite{fauque2013magnetothermoelectric}. Yet this practice for $S_{xx}$ seems invalid in CoSi. As shown in Fig.~\ref{f3}(a), $\Delta S_{xx}$ and $\Delta S_{xy}$ are anti-phase for the mean-frequency below 10 T, and the beating patterns are in-phase. Above 11 T, the mean-frequency changes to in-phase, but the phase shift of the beating patterns cannot be resolved.

To solve this paradox we go back to the definition of $\bm{S}$. In an infinite medium, the total current density in the presence of a temperature gradient and magnetic field can be formulated~\cite{ziman2001electrons} as
$$\bm{J}=\bm\sigma\cdot\bm{E}-\bm{\epsilon}\cdot\nabla T$$
where $\bm{\epsilon}$ is the thermoelectric tensor. By definition, $S_{ij}=E_i/\nabla_jT$, and $\bm{S}=\bm{\rho}\cdot\bm{\epsilon}$. For the oscillatory part (denoted by the upper tilde) in $\bm{\epsilon}$ , Fletcher \it et al.\rm \ have given a simple formula~\cite{fletcher1981amplitude,fletcher1983experimental,coleridge1989low,fletcher1995oscillations}
$$\bm{\widetilde{\epsilon}}=\mp i \frac{\pi k_B}{e}\frac{D'(\mathbb{X})}{D(\mathbb{X})}\bm{\widetilde{\sigma}}=\beta\bm{\widetilde{\sigma}}$$
with the upper sign (negative) for electrons (N. B. here we neglect the contribution from non-diffusive part again, and $\bm{\widetilde{\sigma}}$ is not necessarily observable). Note the negative sign in $e$ and $D'(\mathbb{X})$, and $\beta=\mp i|\beta|$ for electrons and holes, respectively. It's easy to get the form
$$\bm{\widetilde{S}}\simeq \beta \bm{\bar\rho}\cdot\bm{\widetilde{\sigma}}$$
with $\bm{\bar\rho}$ denotes the non-oscillatory part of resistivity tensor. For the metals where $\bar\rho_{xx}\gg\bar\rho_{yx}$, e.g. ZrSiS~\cite{matusiak2017thermoelectric}, it's safe to use $\widetilde{S}_{xx}\simeq \beta\bar\rho_{xx}\widetilde{\sigma}_{xx}$. However $\bar\rho_{xx}$ and $\bar\rho_{yx}$ are of the same magnitude in our CoSi sample, therefore the diagonal and off-diagonal parts get entangled in $S_{ij}$. Remember $\bm{\widetilde{\sigma}}$ is linked with the oscillatory part of density of state $\widetilde{g}$~\cite{coleridge1989low}
$$\widetilde{\sigma}_{xx}\propto\frac{\sigma_0}{1+\omega_c^2\tau_0^2}\frac{\widetilde{g}}{g_0},\quad \widetilde{\sigma}_{xy}\propto\frac{\sigma_0}{(1+\omega_c^2\tau_0^2)\omega_c\tau_0}\frac{\widetilde{g}}{g_0}$$
where $\omega_c$ is the cyclotron frequency $|e|B/m^\ast$, $\tau$ is the quantum life time and the subscript 0 denotes the value at zero field.
For electron-like bands, $\widetilde{\sigma}_{xx}$ and $\widetilde{\sigma}_{xy}$ are in-phase. As $\widetilde{S}_{xy}$ approximately equals $\beta (\bar\rho_{xx}\widetilde{\sigma}_{xy}-\bar\rho_{yx}\widetilde{\sigma}_{xx})$, $\widetilde{S}_{xy}$, $\widetilde{\epsilon}_{xx}$ and $\widetilde{\epsilon}_{xy}$ should be in-phase and have an additional $-i$ phase with respect to $\widetilde{g}$. As $\widetilde{S}_{xx}$ approximates $\beta (\bar\rho_{xx}\widetilde{\sigma}_{xx}+\bar\rho_{yx}\widetilde{\sigma}_{xy})$, the phase is actually dependent on the relative magnitude of $\bar\rho_{xx}\widetilde{\sigma}_{xx}$ and $-\bar\rho_{yx}\widetilde{\sigma}_{xy}$. For holes, similar results are expected, except for that $\widetilde{\epsilon}_{xx}$ and $\widetilde{\epsilon}_{xy}$ have a $+i$ phase with respect to $\widetilde{g}$ while $\widetilde{S}_{xy}$ has an additional $-i$ phase.

We plot the oscillatory parts of $\epsilon_{ij}$ in Fig.~\ref{f3}(b), in comparison with the oscillatory parts of $S_{ij}$ in Fig.~\ref{f3}(a). Apparently the phases of $\widetilde{S}_{xy}$, $\widetilde{\epsilon}_{xx}$ and $\widetilde{\epsilon}_{xy}$ are perfectly in-phase, which again indicates that these oscillations are stemming from electron-like pockets.

Because there exist only two distinct frequencies in QOs, we are able to separate them thoroughly using FFT filter, as shown in Fig.~\ref{f3}(c) for $\Delta\epsilon_{xy}$. No signs of Zeeman splitting are found, due to the small SOC strength and limited field range of our experiments. The peak positions are assigned with integral Landau indices while the valley positions with half-integrals. Using the so-called Landau fan diagram (Fig.~\ref{f3}(d)), we extract a phase of 0.11 and 0.16 for $\alpha$ and $\beta$, respectively. The total phase shift $\phi_s$ equals $-1/2+\phi_B+\phi_{3D}+\phi_T$, where $\phi_B$ is the Berry phase, $\phi_T$ is $-1/4$ (namely $-i$) here for electrons in thermoelectric QOs, and the additional phase shift $\phi_{3D}$ stemming from the dispersion along $k_z$ equals $-1/8$ for a maximum cross section of electrons~\cite{li2018rules}. We finally get the Berry phases for $\alpha$ and $\beta$ as -0.015 and 0.035, respectively. The angle-dependence of the Berry phases is also obtained from the Nernst voltage measurement when the field is along different orientations.
As shown in Fig.~\ref{f3}(e), the Berry phases are close to zero over the whole $2\pi$ angle for these two pockets.

\begin{figure}[htbp]
\begin{center}
\includegraphics[clip, width=0.49\textwidth]{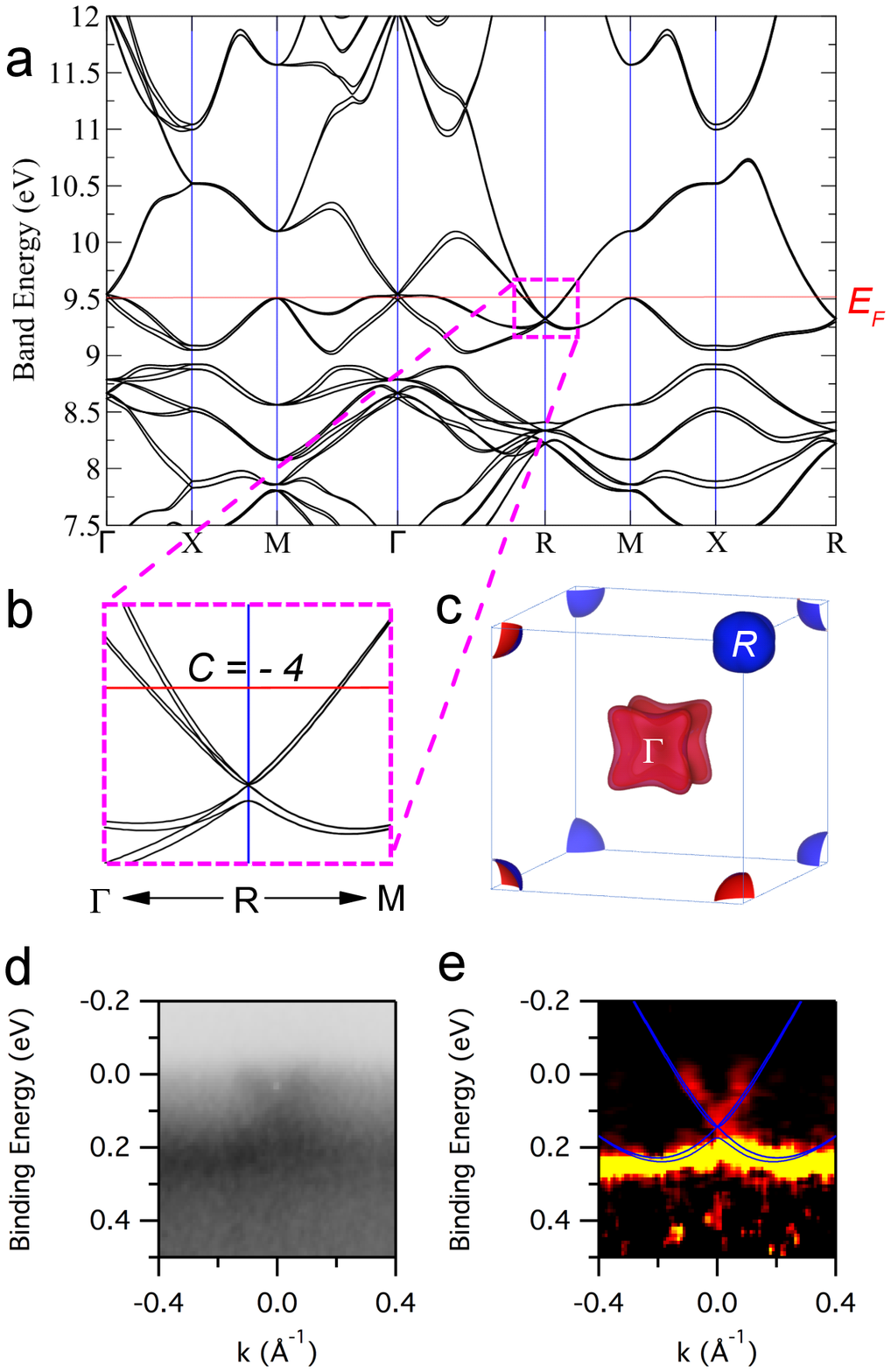}\\[1pt]  
\caption{Color online. (a) Band structure of CoSi when SOC is included.   (b) Zoom-in around $R$ point. (c) Calculated Fermi surfaces of CoSi. (d) ARPES spectra and (e) curvature plot of double Weyl fermion at the $R$ point. Blue lines are band calculation results with 30~meV downward shift of Fermi energy.}
\label{f4}
\end{center}
\end{figure}

To shed light on the electron pockets evolving these QOs, we check the DFT calculated band structure of CoSi in Fig.~\ref{f4}(a). There are electron-like pockets centered at $R$ point in BZ corner and hole-like pockets at $\Gamma$ point in BZ center.
The hole pockets around $\Gamma$ stem from rather flat band and are too large and heavy for transport experiments to detect.
At $R$ point, when SOC is neglected, a band crossing with fourfold degeneracy below the Fermi level occurs. This is a double Weyl fermion with the Chern number $-2$~\cite{PhysRevLett206401,PhysRevLett206402,pshenay2018band}.
When SOC is included, a chiral double sixfold-degenerate spin-1 Weyl node carrying a $-4$ topological charge is formed at $R$ point. This Weyl node lies around 180~meV below Fermi level and it hosts Fermi surfaces of two pairs of near-spherical concentric sheets with different spin textures but similar size.
As Co and Si are light elements, the SOC strength is actually small and therefore the band-splitting is of the order of 10~meV near the Fermi level (band-splitting is around 15~meV along $\Gamma-R$ direction near the Fermi level and the bands become twofold-degenerated along $k_{x,y,z}=\pi$~\cite{PhysRevLett206401}).
Our ARPES experiment demonstrates the double Weyl fermion at $R$ point, but is unable to resolve the $\alpha$ and $\beta$ branches due to instrumental energy resolution (Fig.~\ref{f4}(d) and (e)). This result is consistent with previous observation in CoSi and RhSi of a $\mp 2$  Chern number at $R$ and $\Gamma$ points, respectively~\cite{sanchez2018discovery,PhysRevLett.122.076402,rao2019new}.
Our DFT calculation shows that each pair of Fermi surfaces should have 10\% difference in the extremal cross-section areas within the (111) plane due to SOC-induced splitting.
However we only observed two spin-degenerated orbits ($\alpha$ and $\beta$) in the QOs, and their zero Berry phases are consistent with the $-2$ Chern number of $R$ point.

\begin{figure}[htbp]
\begin{center}
\includegraphics[clip, width=0.5\textwidth]{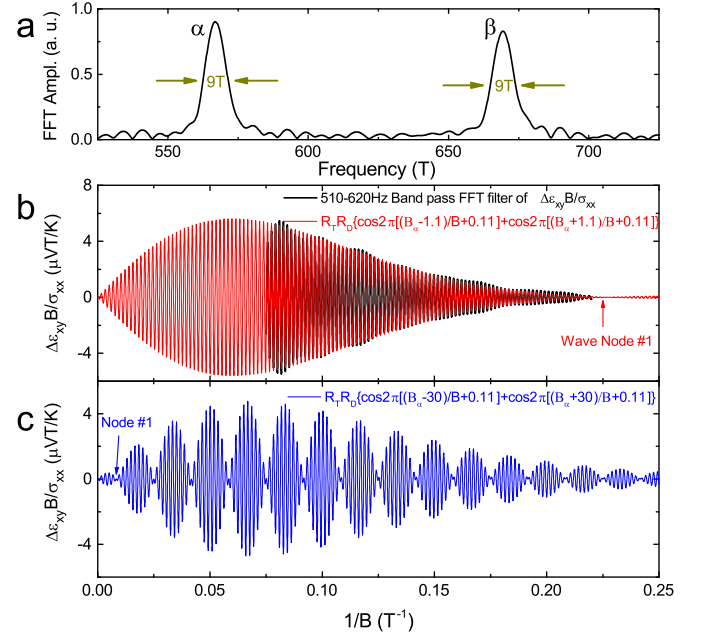}\\[1pt]  
\caption{Color online. (a) FFT spectrum of $\Delta\epsilon_{xy}$. The full widths at half maxima for both $\alpha$ and $\beta$ pockets are around 9~T.  (b) $\Delta \epsilon_{xy}B/\sigma_{xx}$ after passing a 510-620 Hz band FFT filter. The coefficient $B/\sigma_{xx}$ is used for the conformity to the LK formula.
The red line is a simulation using two frequencies close to $B_\alpha$ (567~T and 569~T). (c) A simulation using two frequencies estimated from DFT calculation ($B_\alpha\pm 30\mathrm{T}$) in comparison to (b).}
\label{f5}
\end{center}
\end{figure}

Our measured oscillatory frequencies (568~T and 671~T) match the DFT calculation without SOC (574~T and 660~T) very well if the Fermi level is tuned down for about 30~meV.
The Fermi energy $E_F$ of the electron pockets is estimated via $E_F=\hbar^2k_F^2/m^\ast$, to be averaged around $160-140$ meV (using $m^\ast$ from $S_{xy}$), which is consistent with the DFT calculation.
We also estimated the carrier concentrations for electrons from the QOs by using the equation $n_{QO}=k_F^3/6\pi^2$. For the $\alpha$ and $\beta$ pockets, the density is $3.82\times10^{19}\mathrm{cm^{-3}}$ and $4.90\times10^{19}\mathrm{cm^{-3}}$, respectively. Considering the spin-degeneracy, the total electron density is estimated as $1.74\times10^{20}\mathrm{cm^{-3}}$, close to the value obtained from two-band model ($1.47\times10^{20}\mathrm{cm^{-3}}$ for electrons).

We now estimate the upper limit of the SOC-induced band splitting in our measurements.
As shown in Fig.~\ref{f5}(a), the full widths at half maxima for both $\alpha$ and $\beta$ pockets are around 9~T, which sets an upper limit of the difference between the SOC-split orbits.
To address a more acute upper limit, we consider the detailed profile of the FFT filtered oscillation (Fig.~\ref{f5}(b)).
If we assume that the $B_\alpha$ is composed of two close frequencies similar in magnitude, the interference of the two oscillations will create wave nodes in the oscillatory spectrum.
As there is no phase-inversion in the band-pass FFT filter in the field range of our measurement, we conclude that the first wave node appears at least after $0.22\,\mathrm{T}^{-1}$ (Fig.~\ref{f5}(b)). This gives the limitation of less than 2.2~T for the separation of the two frequencies induced by SOC. By comparison, the DFT with SOC expects a difference of around 60~T in the (111) plane, which will generate a more complicated pattern with multiple wave nodes within the observable field range (Fig.~\ref{f5}(c)). Same results hold for $\beta$ orbit.
Our analysis reveal that the SOC strength at the Fermi level is one order of magnitude smaller than what is expected by DFT. The SOC-induced energy split along $\Gamma-R$ near the Fermi level should be less than 2~meV even when thermal broadening and Zeeman splitting are considered.

\begin{figure}[htbp]
\begin{center}
\includegraphics[clip, width=0.5\textwidth]{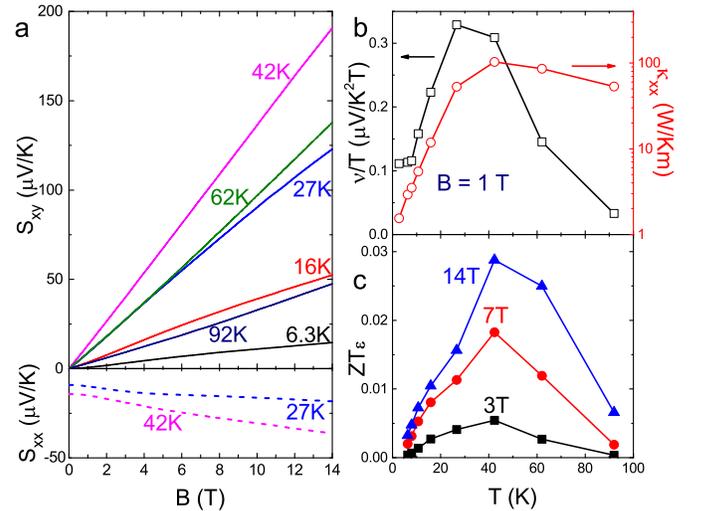}\\[1pt]  
\caption{Color online. (a) $S_{xy}$ and  $S_{xx}$ of T02 at higher temperatures. (b) Nernst coefficient divided by temperature ($\nu/T$) and thermal conductivity ($\kappa_{xx}$) at 1 T. (c) Nernst-Ettingshausen figure of merit (ZT$_\epsilon$) under different magnetic fields.}
\label{f6}
\end{center}
\end{figure}

Lastly, we consider the phonon-drag effect in CoSi at intermediate temperatures.
As shown in Fig.~\ref{f6}(a), $S_{xy}$ has a linear dependence on magnetic field over 6~ K to 100~K and shows no sign of saturation at 14~T.
At 42~K, $S_{xy}$ attains a maximum of around 190~$\mathrm{\mu V/K}$ in 14~T. By contrast, $S_{xx}$ is much smaller under the same magnetic field. A large $S_{xy}$ is usually explained as a resultant of a small Fermi energy and large carrier mobility~\cite{behnia2016nernst,behnia2007nernst}.
In low field limit, the diffusive Nernst coefficient $\nu$ for electrons can be formulated~\cite{tieke1996magnetothermoelectric,behnia2016nernst,behnia2007nernst} as
$$\nu/T=S_{xy}/BT=-\frac{\pi^2}{3}\frac{k_B}{e}\frac{k_B}{\epsilon_F}p\mu$$
where $p=(\frac{\partial ln\tau}{\partial ln \epsilon})|_{\epsilon_F}$, a value usually close to unity~\cite{behnia2015fundamentals}. Using the mobility and Fermi energy from our transport data, we have $\nu/T\sim0.1p\ \mathrm{\mu V/K^2T}$ at 6 K, which is quite close to our observation in Fig.~\ref{f5}(b). The $\nu/T$ increases rapidly with increasing temperature and peaks around 40~K, which suggests an additional large phonon-drag contribution at intermediate temperatures. Although thermal conductivity also peaks at the same region, a large ZT$_\epsilon\!=\!S_{xy}^2T/\rho\kappa\!=\!0.03$ is achieved at 42~K in 14~T. This value surpasses the ordinary ZT of CoSi at room temperature and is comparable to those of hole-doped CoSi samples~\cite{lue2004hole}.
Despite the necessity of a strong magnetic field, this points to possible thermoelectric application of CoSi below liquid nitrogen temperature.

\section{Conclusion}
We report high-quality single-crystal growth of the topological chiral semimetal CoSi via tellurium flux.
These single crystals have large carrier mobilities, which enables us to observe QOs in its thermoelectrical signals.
The oscillatory frequencies reveal two spherical Fermi surfaces around the BZ corner $R$ point, and our measurements suggest that the SOC strength in CoSi is one order of magnitude smaller than DFT calculation.
The extracted Berry phases of these electron orbits agree well with the scenario of $-2$ chiral charge at $R$.
Besides we report a relative large ZT$_\epsilon$ of 0.03 in CoSi at 42~K in 14~T, due to the high mobility and phonon-drag contribution.
Our tellurium flux method might be feasible in growing high-quality crystals of other transition-metal silicide, like the Kondo insulator FeSi and the helimagnet MnSi.
Our findings also highlight the potentials of magneto-thermoelectric measurement for detecting high-frequency QOs in topological semimetals~\cite{PhysRevB.99.104516}.

\section{Acknowledgement}
This work was supported by the National Natural Science Foundation of China No. U1832214, No.11774007, the National Key R\&D Program of China (2018YFA0305601) and the Strategic Priority Research Program of Chinese Academy of Sciences (Grant No. XDB28000000). T.-R.C. was supported from Young Scholar Fellowship Program by Ministry of Science and Technology (MOST) in Taiwan, under MOST Grant for the Columbus Program MOST108-2636-M-006-002, National Cheng Kung University, Taiwan, and National Center for Theoretical Sciences (NCTS), Taiwan. This work is supported partially by the MOST, Taiwan, Grants No. MOST 107-2627-E-006-001. Work at Princeton and Princeton-led ARPES measurements were supported by the US Department of Energy under Basic Energy Sciences (Grant No. DOE/BES DE-FG-02-05ER46200). M.Z.H. acknowledges Visiting Scientist support from Lawrence Berkeley National Laboratory. The authors thank V. Strokov for support at the ADRESS beamline of the Swiss Light Source at the Paul Scherrer Institute in Switzerland. W.X. at LSU was supported by a Beckman Young Investigator award. X.G. at LSU is supported by the National Science Foundation under NSF-OIA-1832967.

%

%

\clearpage

\end{document}